# Relevance Feedback in Conceptual Image Retrieval: A User Evaluation

José Torres [1], Luís Paulo Reis [2,3]

jtorres@ufp.pt, lpreis@fe.up.pt

[1] University Fernando Pessoa, Praça 9 de Abril, 349, 4249-004 Porto, Portugal
[2] LIACC/FEUP – Artificial Intelligence and Computer Science Lab., Faculty of Engineering of the University of Porto, Rua Dr. Roberto Frias, 4200-465 Porto, Portugal

**Abstract:** The Visual Object Information Retrieval (VOIR) system described in this paper implements an image retrieval approach that combines two layers, the conceptual and the visual layer. It uses terms from a textual thesaurus to represent the conceptual information and also works with image regions, the visual information. The terms are related with the image regions through a weighted association enabling the execution of concept-level queries. VOIR uses region-based relevance feedback to improve the quality of the results in each query session and to discover new associations between text and image. This paper describes a user-centred and task-oriented comparative evaluation of VOIR which was undertaken considering three distinct versions of VOIR: a full-fledge version; one supporting relevance feedback only at image level; and a third version not supporting relevance feedback at all. The evaluation performed showed the usefulness of region based relevance feedback in the context of VOIR prototype.

## 1. Introduction

An image retrieval system is a computer-based system for browsing, searching and retrieving images from large databases containing digital images. Content-based image retrieval (CBIR) is the application of computer vision to the image retrieval problem. The search makes use of the contents of the images themselves, rather than relying only on human-inputted metadata such as captions or keywords. The ideal CBIR system, from a user perspective, would involve conceptual retrieval. The user would be able to perform a request such as "find pictures of fishes". This type of query is very difficult for computers because there are all types of fishes of different sizes and shapes and other animals such as, for instance, dolphins that resemble a lot fishes.

The main objective of a CBIR system is the satisfaction of the user needs for some type of visual information. The design and conception of an image retrieval system should, consequently, follow the guidelines offered by the correct observation of what the users really want from the system. In practice, there are three fundamental aspects to be taken into account that make this task difficult:
- The diversity of applications for digital images;
- The diversity of image users with different perspectives, making the problem of requirement definition extremely complex;
- The limitation, within current state of the art, of science and technology to mimic the human capacity of image understanding and description.

A key requirement for developing future image retrieval systems is to explore the synergy between humans and computers. Relevance feedback (RF) is a technique that engages the user and the retrieval system in a process of symbiosis. Following the formulation of the initial query, for subsequent iterations of query refinement, the system presents a set of results and the user evaluates the results in order to refine the set of images retrieved to his or her satisfaction. In image retrieval systems, this technique can be extremely useful to reduce the adverse effects of the three aspects mentioned above.

This paper analyses the use of relevance feedback in image retrieval, presenting VOIR (Torres, 2005) a prototype image retrieval system. The paper also describes a user-centred and task-oriented comparative evaluation of VOIR which was undertaken considering three distinct versions of VOIR: a full-fledge version; one supporting relevance feedback only at image level; and a third version not supporting relevance feedback at all.

The rest of this paper is organised as follows. Section 2 reviews some of the work done regarding user needs and image description in image retrieval. Section 3 introduces the use of relevance feedback. In section 4 details of the VOIR framework are presented in addition to the two-layer model of description of visual items. Section 5 and 6 presents the methodology and the experimental results obtained in the user study carried out. The final section gives some concluding remarks.

## 2. User Needs and Image Description in Image Retrieval

### 2.1. User Needs in Image Retrieval

The purpose for which the images are required typically determines user needs and behaviour when searching for images. It is widely accepted that present day society is much more dependent on the use of visual information in both forms: still and moving images. Visual information is useless if it cannot be obtained in an efficient and effective way. It is of recognised importance that the user needs should be an important part of the requirements used to develop image retrieval systems.

Since the first quarter of the 20th-century, developments in photography led to the widespread use of photograph in the worldwide press. Subsequently, several institutions were concerned with archiving visual material in order to support services to individuals or organisations. These services presuppose that the material is available and easily reachable.

Image archives constitute a natural place to develop user studies in practices of retrieving images. Several studies have focused on user needs in image repositories such as Armitage & Enser (1997), Fidel (1997), Jorgensen (1996) and Ornager (1997).

As for textual documents, one can state that nowadays it is easy to generate visual documents, not so easy to gain physical access to them, and even more difficult to retrieve or access those few visual documents which satisfy a specific information need (Enser, 1995).

In designing an image retrieval system, a crucial aspect is to predict typical behaviour patterns of the users, and identify their needs and expectations. In order to do this, the first task is to identify and classify the different categories of image users, not only the users that depend on the use of images in their professional activity but also those who deal with images for entertainment or recreational purposes.

To provide a full description of visual information applications in diverse domains is extremely difficult. The following categories are not exhaustive but could be interpreted as a description of some of the most representative professional activity types that, in some way, depend on the use of images: Medicine; Crime prevention; Fashion and graphic design; Advertising; Architectural and engineering design; Historical research; Education; Publishing industry and the press.

If it can be reliably established that different types of users do in fact require different styles of interaction with image retrieval systems, the task of the systems designer will be made considerably easier (Eakins & Graham, 1999).

## 2.2. Conceptual Description of Images

The aim of visual information description is to transform user needs into a suitable form to support searching in visual collections. The selected image indexing attributes should be sufficiently discriminatory to allow images to be retrieved in an effective and efficient way.

Ideally, in an image retrieval system, the descriptive information associated with the images should be closely related with the way end users, i.e., the humans, interpret those images.

In the description of fine art pieces, Panofsky (1970), an art historian particularly interested in the analysis of visual art, identifies three levels of image analysis:
- Pre-iconographic: this first level deals with the description of the image motifs such as objects and events. It refers to the essentially factual and expressional facets of the image;
- Iconographic: expressing secondary subject matters such as image interpretations. It presumes that the agent describing the image is familiar with specific themes as transmitted by literary sources;
- Iconology: the third level captures intrinsic meaning of the image and involves association with symbolic values or trends of the human mind.

Using advertising images as examples, Barthes (1977), a social and literary critic with well-known published work about the study of signs and signification, established a semiological theory that extends also to other pictorial forms of expression. In his theory, he distinguishes two different levels of image analyses: denotation and connotation. Denotation may be viewed as a neutral expression of the visual signs, although these are the result of the meaning assigned by a given system or language within a culture. The second level is expressed as connotative meanings relating to feelings, associations and aesthetics considerations. Barthes identifies that the photographs considered convey three types of message:

- Linguistic: the textual message annexed to the photograph, if any;
- Literal: a descriptive or denotative message that identifies the objects in the photograph, i.e., the facts;
- Symbolic: part of the connotation level, and reflects the subjectivity of the photograph, depending on individual or cultural experience and knowledge.

It is interesting to observe that both, Panofsky and Barthes, agree in the fact that the analysis of one particular image gives origin to a part of the description that is objective or factual and other part that depends on the interpreter agent, i.e., is subjective in nature. This supposition is inherently important for the definition of the descriptive information that one generic image retrieval system should support.

## 3. Relevance Feedback in Image Retrieval

*Relevance Feedback* constitutes the process of refining the results returned by the CBIR system in a given iteration of an interaction session. The user performs some sort of evaluation over the results returned in the last iteration and this evaluation is fed back to the system (Figure 1).

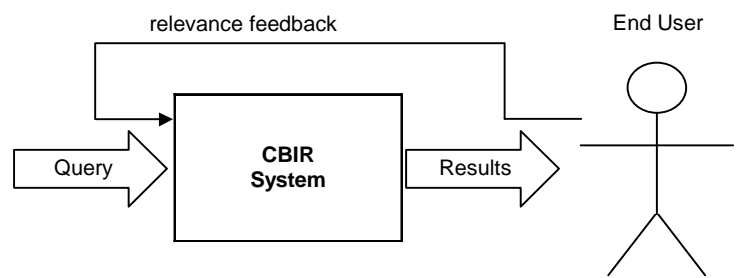

Figure 1 – Relevance Feedback in CBIR

The refinement is possible since the CBIR relates this information with the information from the original query and from other refinements in previous iterations. According to Croft (1995) the process of relevance feedback is one of the preferred characteristics mentioned by users of information retrieval systems.

The two more popular approaches for relevance feedback presented below are classified in Ishikawa et al. (1998) as *query-point movement* and *re-weighting*. These techniques, which aim for the convergence of the query with user needs after relevance feedback iteration, rely on the assumption that the feature space uses a vector space model. In this model, a point in the multidimensional feature space can be used to define each query and the same applies to each item in the database. Also, the user has the opportunity to optionally evaluate each document as having positive or negative relevance, or to simply not evaluate. The two methods are:

- Query-point movement: this method attempts to move the query-point towards a supposed "ideal query-point". This is done by moving the query-point towards the positive examples fed back by the user and away from the negative ones (Ishikawa, Subramanya, & Faloutsos, 1998). Rocchio's formula (Rocchio, 1971), presented in Harman (1992) as the *Standard Rocchio* formula, is the formalization of this strategy on a vector space model:

$$Q_{k+1} = Q_k + \beta . \sum_{i=1}^{n_1} \frac{R_i}{n_1} - \gamma . \sum_{j=1}^{n_2} \frac{S_j}{n_2} \qquad (1)$$

The new query vector, $Q_{k+1}$, is basically a weighted average where $Q_k$ is the query vector descriptor for the previous query iteration $k$, $R_i$ is the vector descriptor for the relevant document $i$, $S_j$ is the vector descriptor for the non-relevant document $j$, $n_1$ is the number of documents selected as relevant by the user, and $n_2$ the number of documents selected as non-relevant. It maximizes the difference between the average score of relevant documents and the average score of non-relevant documents. Finally $\beta$ and $\gamma$ are appropriate positive constants. This model is described by Rui et al. (1997) and has been implemented in one early version of the MARS multimedia retrieval system.

- Re-weighting: this method assumes that there are several features and each feature have its similarity measure $S_i$ producing a list of scores for a given query point. The final score between a query $q$ and an object $o$ is:

$$S(q,o) = \sum_{i=0}^{M-1} W_i \times S_i(q,o) \qquad (2)$$

Where $W_i$ is the weight of similarity measure $S_i$ in a model having M similarity measures. This method works in the similarity measure vector space, re-weighting $W_i$ according to the importance of similarity measure $S_i$ (inter-weight). Consider that feature $i$ is represented by a vector having K components and that each component $j$ has a weight $w_j$. The re-weighting method also applies to those weights (intra-weight). This technique was used in the version of the MARS system described by Rui (1999) and by Rui et al. (1998). In it, the weight for the *j*-th feature component is the value $w_j = 1/\sigma_j$, where $\sigma_j$ is the standard deviation of the good examples along the *j*-th feature component. This reflects that, for example, a feature component with low standard deviation in the set of good examples provided by the user, tends to be more important in the calculation of the similarity measure value since it has more discriminate power. Logically, high standard deviation values mean that the feature will lose importance in the similarity measure.

## 4. VOIR Retrieval Framework

### 4.1. Conceptual Image Retrieval Framework

The VOIR framework aims to be used in conceptual image retrieval. It assumes that the target images of the user are fundamentally associated with concepts, such as, cars, chairs or airplanes. Each concept is represented by a textual term from a textual thesaurus, i.e., a hierarchic controlled vocabulary.

A region-based approach is used for representation, query and retrieval of images. It is assumed that the images were already segmented into regions before being indexed. During the indexing operation, each region is uniquely associated with a feature vector, $f_i$, representing low-level features such as colour, texture and shape. During query formulation, the user chooses textual terms from the thesaurus representing the desired concepts, and then selects, for each term, one of the visual regions already associated with the term to be used as the example during the content-based query.

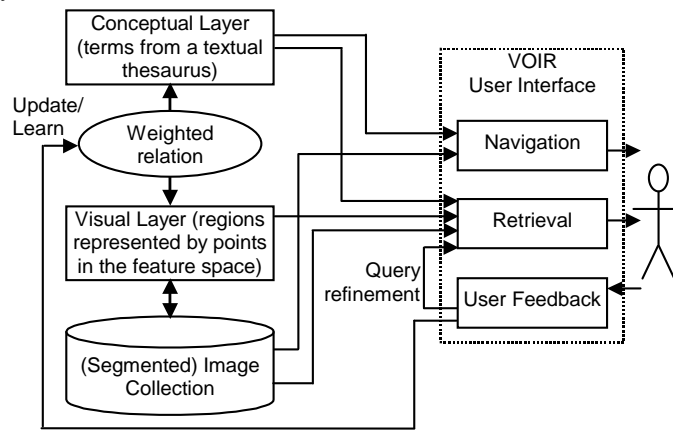

Figure 2 – Overview of the framework for conceptual image retrieval

Low-level features and conventional distance functions, usually, are not sufficient to support the correct discrimination of conceptual similarity between distinct visual regions.

VOIR framework implements a two-layer model separating conceptual categories at the upper layer from the visual layer composed by the low-level feature points. The visual layer is partitioned into visual categories, $V_j$. Each conceptual category, $C_i$, can be related with several visual categories. Each visual category is composed of several regions. The regions sharing the same visual category are conceptually and visually similar. The use of a textual thesaurus reduces inconsistency in term assignment and provides a knowledge structure that can be explored during the searching process.

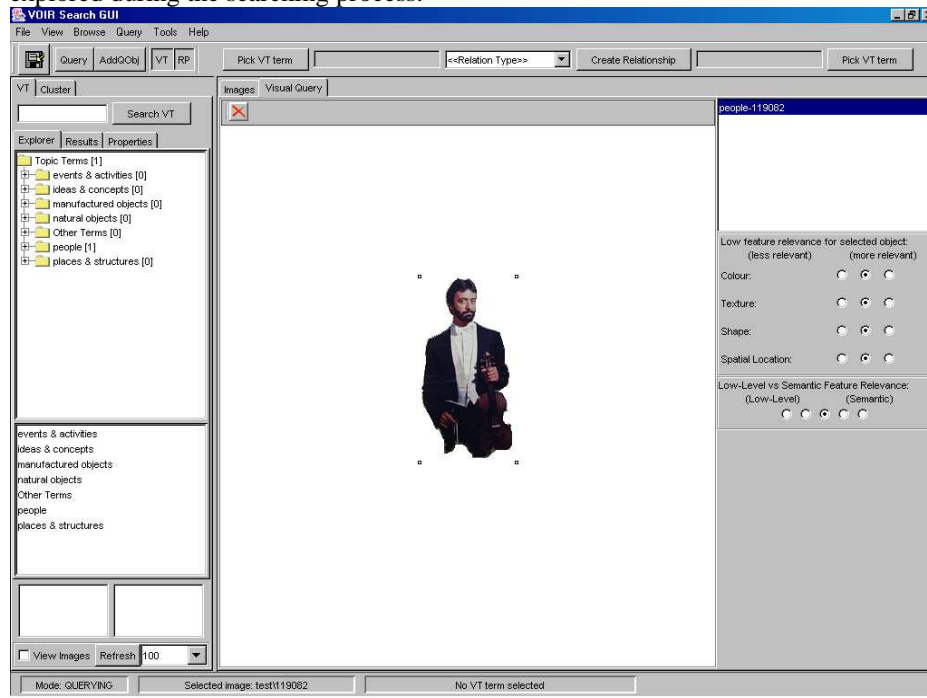

Figure 3 – Snapshot of VOIR GUI interface for query composition

### 4.2. Region-based Relevance Feedback

The region-based relevance feedback information provided by the user supports refinement of the results and, additionally, is used to improve the behaviour of the image retrieval system in subsequent sessions. In the latter situation, the system is said to be evolving over time since it is learning the correct associations between terms and regions.

In each query session, the system implements a relevance feedback mechanism that attempts to move the query point towards the good points and away from the bad points. It also attempts to reweigh the query so as to increase the weight of the more discriminating features. These two methods have been used elsewhere (Rui, Huang, Ortega, & Mehrotra, 1998). The novelty of our approach is that, instead of limiting the number of query points to just one, it can expand the query by using additional query points in the feature space that are related with the same conceptual category.

When a new relevant example $f_i$ is indicated by the user, a Boolean function will indicate if the designated point belongs to the same visual category of the evaluated visual item $f_j$ or not. If true, the new point will be considered as one more positive point of the evaluated item. If false, this point will be considered as the seed of another visual category to be added to the current query.

The current implementation of the mentioned function, essentially compares distances $D_{ji} = distance(f_j, f_i)$ and $D_{jk} = distance(f_j, f_k)$ where $f_k \in F_K$ the set of all visual items whose category $C_k$ is different of the category $C_i$ of point $f_i$. Basically the query expansion is done if $(D_{ji} / D_{jk}) > thr$, where *thr* is a pre-defined threshold level.

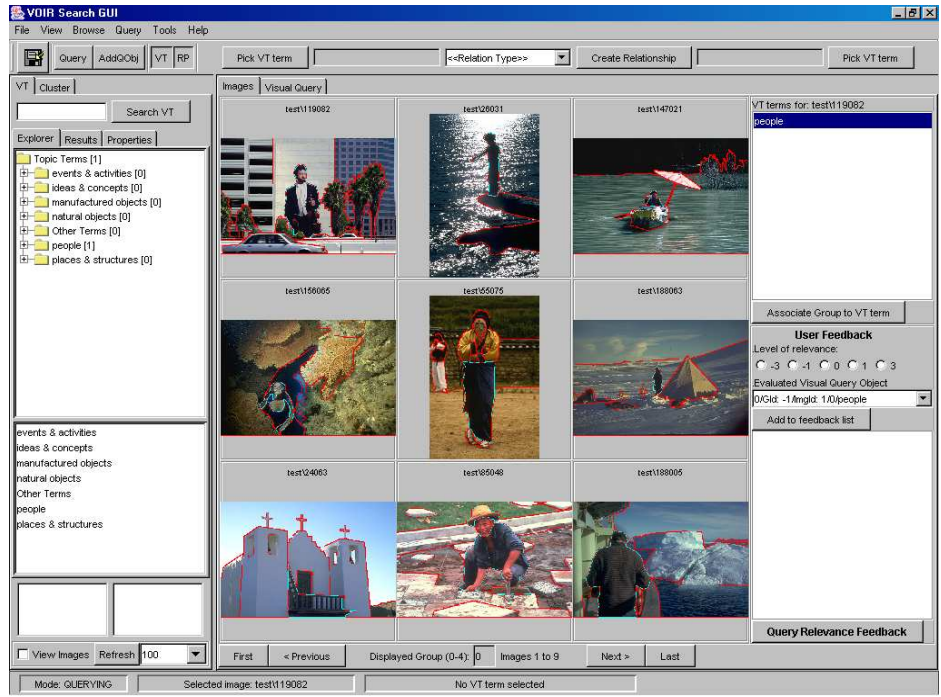

Figure 4 – Snapshot of VOIR GUI interface for result set display and relevance feedback

### 4.3. Learning term-region associations

The association between terms and regions is characterized by having a normalized degree of confidence *d_conf* where the attribute $d\_conf \in [0, 100]$. This association is of fundamental importance since it constitutes the outcome of the process of concept learning. It can be done manually or automatically. In the first case *d_conf* is set to its maximum value (100), in the second case it will be defined or updated algorithmically.

The critical evaluation of the image results by the user during query sessions is used to create or update the existing associations. The outcome of this is that the system gradually learns associations between visual regions and labels from the textual thesaurus. The more the system learns, the more accurate and faster are the subsequent query sessions.

In the implementation used to carry out the experiments, the visual categories, used in the concept learning process, were defined off-line using a clustering algorithm that took low-level features extracted from each region as its input data. The automatic updating of the associations between term and visual item is done periodically after the query sessions or following new manually added associations. The updating process affects all the visual items that belong to the same visual category as the visual item whose situation was changed either because was explicitly associated with a keyword or because was evaluated during a query iteration.

### 4.4. Image Database Used

Although there are actually diverse image datasets annotated in electronic format, virtually all are "per image", i.e., the annotated words are associated with the whole image and the images in the collection are not segmented. This is easily explained by the large manual effort required to the task of annotate the regions on a large segmented image collection. The selected image set is comprised by 300 images. This subset was made publicly available to the research community as part of a project to evaluate segmentation algorithms conducted at University of California, Berkeley (Martin, Fowlkes, Tal, & Malik, 2001). The mentioned dataset belongs to the Corel image database, a large collection of stock photographs widely used in computer vision. The images from this dataset belong to diverse categories such as: animals, plants, people, landscape earth features such as mountains or bushes, manufactured objects such as airplanes. Each image has either 4 or 5 different keywords associated.

## 5. Experimental Methodology

In the carried out study we adopted the within subjects (repeated measures) design and consequently we used just a single set of users, a methodology similar to the one followed by Jose (1998). The independent variable was the VOIR system type. Each member of the set of subjects was asked to interact, on different occasions, with three versions of the system, performing a script of predefined tasks. As outcome, three different sets of results for a range of dependent variables were obtained, which worked as indicative of the user satisfaction, through the administration of a questionnaire per subject.

The three used versions of VOIR system were:
- VOIR-1: didn't supported RF at all. It wasn't capable of discerning, for each image result, in the image ranked result set, which was the best-scored region.
- VOIR-2: supported relevance feedback just at image-level. Didn't supported RF at region-level, neither was capable of discerning for each image result, in the image ranked result set, which was the best-scored region.

- VOIR-3: supported RF at region-level and was capable of discerning for each image result, in the image ranked result set, which was the best-scored region.

The three experimental one-tailed hypotheses tested, in terms of acceptability or degree of satisfaction, were:
- $H_{A1}$: VOIR-2 was more acceptable or satisfying than VOIR-1.
- $H_{A2}$: VOIR-3 was more acceptable or satisfying than VOIR-2.
- $H_{A3}$: VOIR-3 was more acceptable or satisfying than VOIR-1.

The 9 people recruited, as system users for the experiment, were all educators belonging to several degrees of teaching. The idea we had in mind was to put the users accomplishing a task directly related with their professional activities of teachers. We were able to verify, through a pre-search questionnaire, that all our subjects had a good understanding of the task they were assigned. All the subjects have a reasonable command of information technology as end users, and typically use the computer in their professional activities.

We met one subject at a time and on different occasions. For each subject, the procedure sequence carried out is detailed in Figure 5.

1. An introductory orientation session;
2. A pre-search questionnaire;
3. Task 1:
   3.1. A training session on the first system with which the subject was to interact;
   3.2. A hand-out of written instructions for the first task;
   3.3. A search session in which the subject interacted with the first system in pursuit of the first task;
   3.4. A post-search questionnaire;
4. Task 2:
   4.1. A training session on the second system;
   4.2. A hand-out of instructions for the second task;
   4.3. A search session on the second system in pursuit of the second task;
   4.4. A post-search questionnaire;
5. Task 3:
   5.1. A training session on the third system;
   5.2. A hand-out of instructions for the third task;
   5.3. A search session on the second system in pursuit of the third task;
   5.4. A post-search questionnaire;
6. A final questionnaire.

Figure 5 – Procedure Sequence carried out to test VOIR prototype

One of our experimental requirements was that the subjects should be exposed to a simulated work task situation in which their information needs would evolve, in just the same dynamic manner as such needs might be observed to do so in subjects' real working lives.

In the instructions given, each subject was asked to simulate that was creating a leaflet for the promotion of an event, to be held at school, whose generic theme was science/nature. The three imagined events were: "The Tree day", "The World Water day" and "The Birds day". The three leaflets were almost finished and the remaining task to do was the selection of the pictures to illustrate the leaflets. Basically, for each event, the subjects had to select, from an existing image database, 3 images that in their opinion would be appropriate to fill and complete the leaflets being produced. Consequently, for each leaflet, each subject had to perform three searches, one search for each image. During the search process, the user was free to query the database multiple times.

A strategy of counterbalancing was adopted to avoid carryover effects due to exposure to earlier levels in the procedure sequence such as, practice, fatigue or attention. This strategy had determined the order in which each of the 9 subjects interacted with the three versions of VOIR. The established order is the one illustrated in Table 1.

Table 1 – Order of use, by each subject, of each of the three version of VOIR

| Subjects | Version used for task 1 ("The Tree day" leaflet) | Version used for task 2 ("The World Water day" leaflet) | Version used for task 3 ("The Birds day" leaflet) |
|---|---|---|---|
| 1,2,3 | VOIR-1 | VOIR-2 | VOIR-3 |
| 4,5,6 | VOIR-2 | VOIR-3 | VOIR-1 |
| 7,8,9 | VOIR-3 | VOIR-1 | VOIR-2 |

## 6. Experimental Results

### 6.1. Pre-search questionnaire

The pre-search questionnaire was divided in: generic characterization of each subject (gender, age); characterization of the subject's profile in terms of domain of computer technology; specific questions about their typically adopted strategy for search images.

To the question "What is your preferred way for selecting images from a collection", 3 subjects selected the option "Keyword based search system for specifying queries made up of search terms", 2 selected "Unordered sequence of small thumbnail images for browsing through", 4 selected both of the options and none indicated alternating ways not mentioned.

The answers to the open question "What sort of criteria do you use in measuring how successfully you complete a task such as the selection of photographs for the design of a leaflet" were mostly related with features from the content of the image.

### 6.2. Post-search questionnaire

Following an approach similar to Jose et al. (1998) we adopted a two part structure for this questionnaire: (i) a set of semantic differential questions, and (ii) a set of Likert scales questions.

In the semantic differential part, the set of 16, 7-point semantic differential, questions was used to characterize the following four aspects (Table 2):
- First question was dedicated to the task that had been set.
- Two questions focused on the search process carried out by the subject.
- Two focused on the retrieved image set.
- The last 11 questions focused on the system used in the retrieval task.

Table 2 – Post-search questionnarire, with the semantic differentials used

| Was the task: | |
|---|---|
| Clear | Unclear |
| Was the search process: | |
| Interesting | Boring |
| Easy | Difficult |
| Was the retrieval set: | |
| Appropriate | Inappropriate |
| Complete | Incomplete |
| Was the retrieval system: | |
| Efficient | Inefficient |
| Satisfying | Frustrating |
| Reliable | Unreliable |
| Flexible | Rigid |
| Useful | Useless |
| Easy | Difficult |
| Novel | Standard |
| Fast | Slow |
| Simple | Complex |
| Stimulating | Dull |
| Effective | Ineffective |

For each of the 3 possible pairs of system version, (VOIR-1, VOIR-2), (VOIR-1, VOIR-3) and (VOIR-2, VOIR-3), we calculated values of the non-parametric Wilcoxon Matched-Pairs Signed-Ranks (Wilcoxon, 1945). The one-tailed experimental hypotheses, stated before, were tested on the following sets of scores:
- Three tests, one for each of the 3 paired versions, using the combined 99 scores from the semantic differential questions about the retrieval system (11 semantic differentials $\times$ 9 subjects = 99 scores)
- Three tests, one for each of the 3 paired versions, using the combined 144 scores from all the semantic differential questions (16 semantic differentials $\times$ 9 subjects = 144 scores)

Table 3 – Significant level of the one-tailed Wilcoxon Matched-Pairs tests

| One-tailed Hypotheses | Tests using 11 semantic differentials focused on system | Tests using all 16 differentials |
|---|---|---|
| $H_{A1}$ | 0.018 (Null hypothesis rejected) | 0.067 (Null hypothesis not rejected) |
| $H_{A2}$ | 0.022 (Null hypothesis rejected) | 0.00079 (Null hypothesis rejected) |
| $H_{A3}$ | 0.00026 (Null hypothesis rejected) | 0.00003 (Null hypothesis rejected) |

Only in one of the cases, the test of $H_{A1}$ (VOIR-2 more acceptable than VOIR-1) with all the 16 differentials, the null hypothesis couldn't be rejected. Meaningful is the fact that the null hypothesis could be comfortably rejected in all the tests performed when using the differentials directly related with the system version used.

In the 5-point Likert scale questions users were invited to answer, for each statement, their agreement degree, varying from "strongly agree" (1 point) to "strongly disagree" (5 point). Each of the 9 users was asked to answer 3 times, one for each version of the system, to each of the 5 statements in the table. For each 9-score set the mean value was calculated. The Wilcoxon tests were done for each possible pair. Finally, the 9-score, from each of the 5 questions, were combined to form a 45-score set. For each of the three pairs, the Wilcoxon test was calculated. The results obtained are summarised in Table 4.

Table 4 – Means for the five-point Likert scale questions and significant level for the one-tailed Wilcoxon Matched-Pairs tests

| Statement | VOIR-1 9-score mean | VOIR-2 9-score mean | VOIR-3 9-score mean | $H_{A1}$ | $H_{A2}$ | $H_{A3}$ |
|---|---|---|---|---|---|---|
| I had a mental image of a photographs that would satisfy my requirements | 2.11 | 1.67 | 1.11 | 0.094 | 0.031 | 0.016 |
| My query was an accurate representation of the type of image(s) I had in mind | 2.44 | 1.89 | 1.44 | 0.098 | 0.063 | 0.02 |
| I am very happy with the image(s) I chose | 2.11 | 1.78 | 1.22 | 0.16 | 0.031 | 0.0078 |
| I believe that I have seen all the possible photographs that satisfy my requirement | 1.89 | 1.56 | 1.22 | 0.125 | 0.156 | 0.039 |
| I believe that I have succeeded in my performance of the design task | 2.56 | 2.11 | 1.44 | 0.063 | 0.027 | 0.0098 |
| All the 5 statements | 2.22 | 1.80 | 1.29 | 0.0011 | 5.7E-5 | 1.4E-6 |

The global mean for the scores obtained by VOIR-3 (1.29) was significantly less than VOIR-2 (1.80) and this was also less than the one for VOIR-1 (2.22). All the tests confirmed $H_{A3}$. In the fourth statement, the three versions were much closer in terms of score. In none of the 5 statements the null hypotheses for $H_{A1}$ could be rejected, since the level of significance was always above ($p > 0.05$).

### 6.3. Final questionnaire

The final questionnaire was given after each user had completed all his tasks. In this questionnaire, the subjects were asked to rank the three systems in terms of (i) enjoyableness and (ii) helpfulness. Also, they were asked why they chose to rank that way. The results after applying the non-parametric Fisher sign test (Weisstein, 2006) to the three pairs of versions are listed in table 5.

Table 5 – Results of application of the Fisher sign test to the ranked systems

| Characteristic compared | VOIR-2 vs VOIR-1 | | | VOIR-3 vs VOIR-2 | | | VOIR-3 vs VOIR-1 | | |
|---|---|---|---|---|---|---|---|---|---|
| | + | – | $H_{A1}$ | + | – | $H_{A2}$ | + | – | $H_{A3}$ |
| Enjoyableness | 8 | 1 | 0.0196 | 8 | 1 | 0.0196 | 9 | 0 | 0.00196 |
| Helpfulness | 9 | 0 | 0.00196 | 8 | 1 | 0.0196 | 9 | 0 | 0.00196 |

The results achieved show clearly that the users prefer VOIR-3 system finding it clearly better in terms of "Enjoyableness" and "Helpfulness" than its counterparts. The results also showed that users find VOIR-2 more enjoyable and helpful than VOIR-1.

## 7. Conclusions

This paper described a Visual Object Information Retrieval system implementing conceptual image retrieval with two layers: conceptual and visual. VOIR uses region-based relevance feedback to improve the quality of the results in each query session and to discover new associations between text and image.

The system was validated through a user-centred and task-oriented evaluation, comparing it with previous versions without relevance feedback and only with relevance feedback at the image level. The results achieved showed clearly the usefulness of our region based relevance feedback approach.